\documentclass[pra,twocolumn,amssymb]{revtex4}
\usepackage{graphicx}
\usepackage{amsmath}

\begin{document}

\title{Collective excitations of a BCS superfluid in the presence of two sublattices}

\author{M. Iskin}
\affiliation{Department of Physics, Ko\c{c} University, Rumelifeneri Yolu, 
34450 Sar\i yer, Istanbul, Turkey}

\date{\today}

\begin{abstract}

We consider a generic Hamiltonian that is suitable for describing a uniform 
BCS superfluid on a lattice with a two-point basis, and study its collective 
excitations at zero temperature. For this purpose, we first derive 
an effective-Gaussian action for the pairing fluctuations, and then extract 
the low-energy dispersion relations for the in-phase Goldstone and 
out-of-phase Leggett modes along with the corresponding amplitude 
(i.e., the so-called Higgs) ones. We find that while the Goldstone mode is 
gapless at zero momentum and propagating in general, the Leggett 
mode becomes undamped only with sufficiently strong interactions.
Furthermore, we show that, in addition to the conventional contribution 
that is controlled by the energy of the Bloch bands, the velocity of the 
Goldstone mode has a geometric contribution that is governed by the 
quantum metric tensor of the Bloch states. Our results suggest that 
the latter contribution dominates the velocity when the former one 
becomes negligible for a narrow- or a flat-band.

\end{abstract}

\maketitle

\section{Introduction}
\label{sec:intro}

The deeper connection between the quantum geometry of the underlying 
Bloch states and the superfluid (SF) phase stiffness tensor or the often-called 
SF weight of some multi-band Fermi SFs has recently been revealed in
the literature~\cite{peotta15, liang17, iskin18}. It turns out that the SF 
stiffness tensor of a multi-band SF has two physically distinct mechanisms: 
while the conventional contribution is due to the intraband processes 
and has a direct counterpart in the one-band models, the geometric 
contribution is due to the interband processes, and therefore, is exclusive to
the multi-band models. In the particular case of a uniform BCS superfluid 
with two underlying sublattices~\cite{liang17, iskin19a, iskin19b}, this is 
such that the geometric contribution is controlled by the so-called quantum 
metric tensor of the underlying Bloch states~\cite{provost80,berry89,resta11}.

Furthermore, in the context of spin-orbit coupled Fermi SFs in continuum, 
we recently showed that the quantum metric tensor of the underlying 
helicity states has also a partial control over the low-energy collective 
excitations of the system at zero temperature~\cite{iskin19c}. Motivated 
by this result and earlier works~\cite{liang17, iskin19a, iskin19b}, here 
we perform a similar collective-mode analysis to the case of a 
generic Hamiltonian that is suitable for describing a uniform BCS SF 
on a lattice with a two-point basis. Allowing that the SF order parameter 
may fluctuate (around its uniform value) independently on the two
sublattices, there are two phase and two amplitude modes which 
are associated with the total and relative fluctuations of the phase and 
amplitude degrees of freedoms. For instance, the in-phase fluctuations 
are phonon-like and correspond to the Goldstone mode, and the 
out-of-phase fluctuations are exciton-like and correspond to the Leggett 
mode. Thus, in comparison to the Goldstone mode that is considered in 
Ref.~\cite{iskin19c}, the presence of a Leggett mode makes the current 
analysis somewhat more cumbersome. 

We find that while the Goldstone mode is gapless at zero momentum 
and propagating in general, the low-energy Leggett mode becomes 
undamped only with sufficiently strong interactions. More importantly, 
by identifying the quantum-metric contribution to the Goldstone mode, 
we show that this geometric effect is complementary to the recent 
works on the geometric contribution to the SF stiffness 
tensor~\cite{liang17, iskin19a, iskin19b}, i.e., they 
are both controlled by the effective-mass tensor of the SF carriers. 
This suggests that an analogous contribution to the collective excitations 
must be present in many other multi-band systems including 
the twisted bilayer graphene~\cite{hu19, julku19, xie19}.

The rest of the paper is organized as follows. In Sec.~\ref{sec:eaa},
we first derive an effective-Gaussian action for the pairing fluctuations, 
then extract the low-energy dispersion relations for the collective modes,
and then benchmark our generic results with those of the honeycomb
literature~\cite{zhao06, tsuchiya13, zhang17}. In Sec.~\ref{sec:gi}, we 
show that the velocity of the Goldstone mode has a geometric contribution
that can be traced back to the same origin as the recent works on the SF 
stiffness tensor~\cite{liang17, iskin19a, iskin19b}. The paper ends with 
a summary of our conclusions in Sec.~\ref{sec:conc}.

\section{Effective-Action Approach}
\label{sec:eaa}

In this section, we first introduce a generic lattice Hamiltonian that is suitable 
for describing a uniform BCS SF with two underlying sublattices, and then 
extract its collective excitations from an effective action that is derived up 
to the Gaussian order in the fluctuations of the SF order parameter.

\subsection{Hamiltonian}
\label{sec:ham}
Having a general single-particle Hamiltonian on a lattice with a two-point basis 
in mind, we consider
\begin{align}
H &= \sum_{\sigma \mathbf{k}} (c_{\sigma A \mathbf{k}}^\dagger \, c_{\sigma B \mathbf{k}}^\dagger) 
\left[ \xi_{\mathbf{k}} \tau_0  + \mathbf{d}_\mathbf{k} \cdot \boldsymbol{\tau} \right]
\left( \begin{array}{c} c_{\sigma A \mathbf{k}} \\ c_{\sigma B \mathbf{k}} \end{array} \right) \nonumber \\
&- U \sum_{S \mathbf{k} \mathbf{k'} \mathbf{q}} 
c_{\uparrow S, \mathbf{k}}^\dagger 
c_{\downarrow S, -\mathbf{k}+\mathbf{q}}^\dagger
c_{\downarrow S, -\mathbf{k'}+\mathbf{q}}
c_{\uparrow S, \mathbf{k'}},
\label{eqn:ham}
\end{align}
where $c_{\sigma S \mathbf{k}}^\dagger$ ($c_{\sigma S \mathbf{k}}$) creates 
(annihilates) a spin-$\sigma$ fermion on sublattice $S \in \{A, B\}$ with 
quasi-momentum $\mathbf{k}$, i.e., in units of $\hbar \to 1$ the Planck constant. 
In the first line where
$
\xi_\mathbf{k} = \epsilon_\mathbf{k} - \mu
$ 
and $\tau_0$ is a $2 \times 2$ identity matrix, $\epsilon_\mathbf{k}$ is due to 
the intra-sublattice hoppings, and $\mu$ is the chemical potential. 
The inter-sublattice hoppings are taken into account by the second term where
$
\boldsymbol{\tau} = \sum_i \tau_i \boldsymbol{\widehat{i}}
$
is a vector of Pauli matrices for the sublattice sector, and the sublattice-coupling 
field
$
\mathbf{d}_\mathbf{k} = \sum_i d_\mathbf{k}^i \boldsymbol{\widehat{i}}
$
is a generic one with $\boldsymbol{\widehat{i}}$ denoting a unit vector along 
the $i = (x,y,z)$ direction. 
Thus, the single-particle problem is described by the Hamiltonian density
$
h_\mathbf{k}^0 = \epsilon_{\mathbf{k}} \tau_0  + \mathbf{d}_\mathbf{k} \cdot \boldsymbol{\tau},
$
leading to a two-band energy spectrum
$
\epsilon_{s\mathbf{k}} = \epsilon_\mathbf{k} + s d_\mathbf{k}
$ 
with $d_\mathbf{k} = |\mathbf{d}_\mathbf{k}|$, where $s = \pm$ labels 
the upper/lower bands.
For instance, in the case of a honeycomb lattice~\cite{iskin19a}, one finds
$
\epsilon_{\mathbf{k}} = -2t' \cos(\sqrt{3} k_x a) - 4t' \cos(\sqrt{3}k_x a/2) \cos(3k_y a/2),
$
$
d_\mathbf{k}^x = -t \cos(k_y a) - 2t \cos(k_y a/2) \cos(\sqrt{3}k_x a/2),
$
$
d_\mathbf{k}^y = t \sin(k_y a) - 2t \sin(k_y a/2) \cos(\sqrt{3}k_x a/2),
$
and
$
d_\mathbf{k}^z = 0.
$
Similarly, in the case of a Mielke lattice~\cite{iskin19b}, one finds
$
\epsilon_{\mathbf{k}} = -2(t'+t'') \cos(k_x a) \cos(k_y a),
$
$
d_\mathbf{k}^x = -2t \cos(k_x a) - 2t \cos(k_y a),
$
$
d_\mathbf{k}^z = 2(t'-t'') \sin(k_x a) \sin(k_y a),
$
and
$
d_\mathbf{k}^y = 0.
$
Note that while
$
\epsilon_{\mathbf{k}} = \epsilon_{-k_x, k_y} = \epsilon_{k_x, -k_y}
$
and the $\tau_x$ field
$
d_\mathbf{k}^x = d_{-k_x, k_y}^x = d_{k_x, -k_y}^x
$
are parity even functions of both $k_x$ and $k_y$, and the $\tau_z$ field
$
d_\mathbf{k}^z = -d_{-k_x, k_y}^z = -d_{k_x, -k_y}^z
$
is an odd function of both $k_x$ and $k_y$, the $\tau_y$ field
$
d_\mathbf{k}^y = d_{-k_x, k_y}^y = -d_{k_x, -k_y}^y
$
is an even (odd) function of $k_x$ ($k_y$).

In the second line of Eq.~(\ref{eqn:ham}), $U \ge 0$ corresponds to the 
strength of the onsite attraction between $\uparrow$ and $\downarrow$ 
particles, and we decouple this quartic term (in the fermionic degrees of 
freedom) using the Grassmann functional-integral formalism~\cite{iskin05, zhao06}. 
For this purpose, we first express the partition function
$
\mathcal{Z} = \int \mathcal{D}[c^\dagger, c]e^{-\mathcal{S}}
$ 
with the associated action
$
\mathcal{S} = \int_0^{1/T} d\tau [\sum_{\sigma S \mathbf{k}} 
c_{\sigma S \mathbf{k}}^\dagger (\tau) \partial_\tau c_{\sigma S \mathbf{k}} (\tau)
+ H(\tau)],
$
where $T$ is the temperature in units of $k_B \to 1$ the Boltzmann constant.
Then, we introduce a Hubbard-Stratanovich transformation at the expense 
of introducing a complex bosonic field $\Delta_{S q}$, and integrate out 
the remaining terms that are quadratic in the fermionic degrees of freedom. 
This leads to
$
\mathcal{Z} = \int \mathcal{D} [\Delta^*, \Delta] e^{-\mathcal{S}_\mathrm{eff}},
$ 
where $\Delta_{S q}$ plays the role of a fluctuating order parameter for 
pairing, and $\mathcal{S}_\mathrm{eff}$ is the effective bosonic action for 
the resultant pairs of fermions. Here, the collective index 
$q = (\mathbf{q}, \mathrm{i}\nu_n)$ denotes both the pair momentum 
$\mathbf{q}$ and the bosonic Matsubara frequency $\nu_n = 2\pi nT$. 
Finally, by decomposing 
$
\Delta_{S q} = \Delta_0 + \Lambda_{S q}
$ 
in terms of a $q$-independent stationary field $\Delta_0$ and $q$-dependent 
fluctuations around it, one may in principle obtain $\mathcal{S}_\mathrm{eff}$ 
at the desired order in $\Lambda_{S q}$. Note that $\Delta_0$ is uniform 
for the entire lattice. 

In this paper we include only the first nontrivial term and obtain the 
effective-Gaussian action 
$
\mathcal{S}_\mathrm{Gauss} = \mathcal{S}_0 + \mathcal{S}_2,
$ 
as the first-order term $\mathcal{S}_1$ trivially vanishes due to the 
saddle-point condition discussed next.

\subsection{Saddle-point approximation}
\label{sec:spa}

The effective-action approach is a standard tool in many-body 
physics, and it leads to
$
\mathcal{S}_0 = \Delta_0^2/(T U) + [1/(N_lT)] \sum_{s \mathbf{k}} \xi_{s \mathbf{k}}
- (1/N_l) \sum_{k} \ln [\det (\mathbf{G}_k^{-1}/T)],
$
where the collective index $k = (\mathbf{k}, \mathrm{i}\omega_\ell)$ denotes 
both the particle momentum $\mathbf{k}$ and the fermionic Matsubara 
frequency $\omega_\ell = (2\ell + 1)\pi T$. Here, $N_l$ is the number of 
lattice sites, and
$
\mathbf{G}_k^{-1} = \mathrm{i}\omega_\ell \mathbf{1}  - H_\mathbf{k}^0
$
is the inverse Green's function for the mean-field Hamiltonian density 
$H_\mathbf{k}^0$, i.e.,
\begin{align*}
\mathbf{G}_k^{-1} = 
\left[
\begin{array}{cc}
 (\mathrm{i} \omega_\ell - \xi_{\mathbf{k}}) \tau_0 
 - \mathbf{d}_\mathbf{k} \cdot \boldsymbol{\tau} & -\Delta_0 \tau_0 \\
  -\Delta_0 \tau_0 & (\mathrm{i}\omega_\ell + \xi_{\mathbf{k}}) \tau_0 
  + \mathbf{d}_{\mathbf{k}} \cdot \boldsymbol{\tau} \\
\end{array}
\right].
\end{align*}
Here, we use
$
\epsilon_{-\mathbf{k}} = \epsilon_{\mathbf{k}}
$
and
$
\mathbf{d}_{-\mathbf{k}} \cdot \boldsymbol{\tau}^* = \mathbf{d}_{\mathbf{k}} \cdot \boldsymbol{\tau},
$
and the Hamiltonian is given by
$
H_0 = \sum_\mathbf{k} \Psi_\mathbf{k}^\dagger H_\mathbf{k}^0 \Psi_\mathbf{k}
$
where
$
\Psi_\mathbf{k}^\dagger = (c_{\uparrow A \mathbf{k}}^\dagger \, c_{\uparrow B \mathbf{k}}^\dagger
\, c_{\downarrow A, -\mathbf{k}} \, c_{\downarrow B, -\mathbf{k}}).
$
After the summation over $\omega_\ell$, we obtain
\begin{align}
\mathcal{S}_0 = \frac{\Delta_0^2}{T U} 
+ \frac{1}{N_l}\sum_{s \mathbf{k}} \left\lbrace \frac{\xi_{s\mathbf{k}} - E_{s\mathbf{k}}}{T}
+ 2\ln [f(-E_{s\mathbf{k}})] \right\rbrace,
\label{eqn:S0}
\end{align}
where
$
\xi_{s\mathbf{k}} = \epsilon_{s\mathbf{k}} - \mu,
$
$
E_{s\mathbf{k}} = \sqrt{\xi_{s\mathbf{k}}^2+\Delta_0^2}
$
is the quasiparticle energy spectrum, and $f(x) = 1/(e^{x/T} + 1)$ is the 
Fermi-Dirac distribution. 

The saddle-point order parameter $\Delta_0$ can also be expressed as
$
\Delta_0 = U\langle c_{\uparrow S \mathbf{k}} c_{\downarrow S,-\mathbf{k}} \rangle
$
with $\langle \dots \rangle$ denoting a thermal average, and we take it 
to be a real parameter throughout the paper without the loss of generality. 
Using the saddle-point condition $\partial \mathcal{S}_0/\partial \Delta_0 = 0$ for 
the action, and the thermodynamic relation
$
N_0 = - T \partial \mathcal{S}_0 /\partial \mu
$
for the number of particles, we find~\cite{zhao06, iskin19a, iskin19b}
\begin{align}
\label{eqn:op}
\frac{1}{U} &= \frac{1}{N_l} \sum_{s \mathbf{k}} \frac{1 - 2f(E_{s\mathbf{k}})}{2E_{s \mathbf{k}}}, \\
\label{eqn:num}
N_0 &= \sum_{s \mathbf{k}} \bigg\lbrace\frac{1}{2} - \frac{\xi_{s\mathbf{k}}}{2E_{s\mathbf{k}}} 
\left[1 - 2f(E_{s\mathbf{k}}) \right]\bigg\rbrace.
\end{align}
In order to evaluate the collective excitations, we need self-consistent
solutions for $\Delta_0$ and $\mu$ as a function of $U$ and hopping 
parameters. In addition, for the $T = 0$ of interest in this paper, these 
mean-field solutions turns out to be sufficient for a qualitative description
of the many-body problem.

\subsection{Gaussian fluctuations}
\label{sec:gf}

Going beyond the saddle-point action $\mathcal{S}_0$, we calculate the 
first nontrivial term in the expansion, and find
$
\mathcal{S}_2 = \sum_{S q} |\Lambda_{S q}|^2/(2T U)
+ [1/(2N_l)] \mathrm{Tr} \sum_{kq} \mathbf{G}_k \mathbf{\Sigma}_q 
\mathbf{G}_{k+q} \mathbf{\Sigma}_{-q},
$
where $\mathrm{Tr}$ denotes a trace over the sublattice and spin sectors. 
The matrix elements of $\mathbf{G}_k$ can be written as
\begin{align}
\label{eqn:G11}
G_k^{11} &= \frac{1}{2} \sum_s \frac{\mathrm{i} \omega_\ell + \xi_{s \mathbf{k}}}{(\mathrm{i} \omega_\ell)^2 - E_{s\mathbf{k}}^2} 
\left(\tau_0 + s\widehat{\mathbf{d}}_\mathbf{k} \cdot \boldsymbol{\tau} \right), \\
\label{eqn:G22}
G_k^{22}&= \frac{1}{2} \sum_s \frac{\mathrm{i} \omega_\ell - \xi_{s \mathbf{k}}}{(\mathrm{i} \omega_\ell)^2 - E_{s\mathbf{k}}^2} 
\left(\tau_0 + s\widehat{\mathbf{d}}_\mathbf{k} \cdot \boldsymbol{\tau} \right), \\
\label{eqn:G12}
G_k^{12} &= \frac{1}{2} \sum_s \frac{\Delta_0}
{(\mathrm{i} \omega_\ell)^2 - E_{s\mathbf{k}}^2} 
\left(\tau_0 + s\widehat{\mathbf{d}}_\mathbf{k} \cdot \boldsymbol{\tau}\right),
\end{align}
where 
$
\widehat{\mathbf{d}}_\mathbf{k} = \mathbf{d}_\mathbf{k}/d_\mathbf{k},
$
and $G_k^{21} = G_k^{12}$. In addition, the matrix elements of the fluctuation
field $\mathbf{\Sigma}_q$ are
$
\Sigma_q^{11} = \Sigma_q^{22} = 0,
$
$
\Sigma_q^{12} = - \Lambda_{T q} \tau_0 -  \Lambda_{R q} \tau_z,
$
and
$
\Sigma_q^{21} =  - \Lambda_{T,-q}^* \tau_0 -  \Lambda_{R,-q}^* \tau_z.
$
Motivated by the earlier works on two-band SFs, we define
$
\Lambda_{T q} = (\Lambda_{A q} + \Lambda_{B q})/2
$
for the total and
$
\Lambda_{R q} = (\Lambda_{A q} - \Lambda_{B q})/2
$
for the relative fluctuations. 

After the summation over $\omega_\ell$, we obtain
$
\mathcal{S}_2 = [1/(2 N_l T)] \sum_q \mathbf{\bar{\Lambda}}_q^\dagger \mathbf{M}_q \mathbf{\bar{\Lambda}}_q,
$
where
$
\mathbf{\bar{\Lambda}}_q^\dagger = (\Lambda_{T q}^* \, \Lambda_{T, -q} \, \Lambda_{R q}^* \, \Lambda_{R, -q})
$
is a vector of fluctuation fields and 
$
\mathbf{M}_q = 
\left( \begin{array}{c|c}
\mathbf{T}_q & \mathbf{C}_q  \\
\hline
\mathbf{C}_q^* & \mathbf{R}_q 
\end{array} \right)
$ 
stands for the inverse fluctuation propagator. Here, while the submatrices 
$\mathbf{T}_q$ and $\mathbf{R}_q$ describe the purely total and purely 
relative fluctuations, respectively, the submatrix $\mathbf{C}_q$ is 
responsible for their coupling. The submatrix $\mathbf{C}_q^*$ is related
to $\mathbf{C}_q$ via a complex conjugate acting only on the multiplying 
factors, i.e., its matrix elements are determined by Eq.~(\ref{eqn:Cqij}) 
but with $[ d_z + d_z' + \textrm{i} (d_x d_y' - d_y d_x') ]$.

In order to simplify their expressions, we denote 
$\xi_{s \mathbf{k}}$ by $\xi$,
$\xi_{s', \mathbf{k+q}}$ by $\xi'$,
$E_{s \mathbf{k}}$ by $E$,
$E_{s', \mathbf{k+q}}$ by $E'$,
$sd_\mathbf{k}^i/d_\mathbf{k}$ by $d_i$,
$s'd_\mathbf{k+q}^i/d_\mathbf{k+q}$ by $d_i'$,
and define the functions 
$u^2 = (1+\xi/E)/2$,
$u'^2 = (1+\xi'/E')/2$, 
$v^2 = (1-\xi/E)/2$,
$v'^2 = (1-\xi'/E')/2$, 
$f = 1/(e^{E/T}+1)$, and
$f' = 1/(e^{E'/T}+1)$. 
In addition, we also define
\begin{align}
\label{eqn:r1}
r_1 = &(1-f-f') \left( \frac{u^2 u'^2}{\mathrm{i}\nu_n-E-E'} - \frac{v^2 v'^2}{\mathrm{i}\nu_n+E+E'} \right) \nonumber \\
&+ (f-f') \left( \frac{v^2 u'^2}{\mathrm{i}\nu_n+E-E'} - \frac{u^2 v'^2}{\mathrm{i}\nu_n-E+E'} \right), \\
\label{eqn:r2}
r_2 = &(1-f-f') \left( \frac{u v u 'v'}{\mathrm{i}\nu_n+E+E'} - \frac{u v u'v'}{\mathrm{i}\nu_n-E-E'} \right) \nonumber \\
&+ (f-f') \left( \frac{u v u' v'}{\mathrm{i}\nu_n+E-E'} - \frac{u v u' v'}{\mathrm{i}\nu_n-E+E'} \right),
\end{align}
for a compact presentation of the matrix elements of $\mathbf{M}_q$ as well. Using 
these simpler notations and definitions, we find
\begin{align}
T_q^{1j} &= \frac{\delta_{1j}}{U} + \frac{1}{2N_l} \sum_{ss' \mathbf{k}} r_j
( 1 + d_x d_x' + d_y d_y' + d_z d_z' ), \\
R_q^{1j} &= \frac{\delta_{1j}}{U} + \frac{1}{2N_l} \sum_{ss' \mathbf{k}} r_j
(1 - d_x d_x' - d_y d_y' + d_z d_z' ), \\
C_q^{1j} &= \frac{1}{2N_l} \sum_{ss' \mathbf{k}} r_j
[ d_z + d_z' - \textrm{i} (d_x d_y' - d_y d_x') ],
\label{eqn:Cqij}
\end{align}
where $\delta_{ij}$ is the Kronecker-delta~\cite{mfactors}. The remaining 
elements of $\mathbf{T}_q$, $\mathbf{R}_q$ and $\mathbf{C}_q$ are all 
related to the given ones as follows: 
$T_q^{22} = T_{-q}^{11}$, 
$T_q^{21} = T_q^{12}$, 
$R_q^{22} = R_{-q}^{11}$, 
$R_q^{21} = R_q^{12}$, 
$C_q^{22} = C_{-q}^{11*}$ and 
$C_q^{21} = C_q^{12}$.
Here, the complex conjugate again acts only on the multiplying factor 
of Eq.~(\ref{eqn:Cqij}). We note that while $T_q^{12}$ and 
$R_q^{12}$ are even both under $\mathbf{q} \to -\mathbf{q}$ and 
$\mathrm{i}\nu_n \to -\mathrm{i}\nu_n$,  $C_q^{12}$ is even only under 
$\mathrm{i}\nu_n \to -\mathrm{i}\nu_n$, and $T_q^{11}$ and $R_q^{11}$
are even only under $\mathbf{q} \to -\mathbf{q}$.
In addition, we also note that a familiar factor 
$
d_x d_x' + d_y d_y' + d_z d_z' 
= ss' \widehat{\mathbf{d}}_\mathbf{k} \cdot \widehat{\mathbf{d}}_\mathbf{k+q}
$
is appearing in the elements of $\mathbf{T}_q$~\cite{iskin19c}.

Next we reexpress the fluctuation fields
$
\Lambda_{T(R) q} = \alpha_{T(R) q} e^{\mathrm{i}\gamma_{T(R) q}} 
$
in terms of real functions $\alpha_{T(R) q}$ and $\gamma_{T(R) q}$, and associate 
$
\lambda_{T(R) q} = \sqrt{2} \alpha_{T(R) q} \cos(\gamma_{T(R) q})
$
with the amplitude degrees of freedom and
$
\theta_{T(R) q} = \sqrt{2}\alpha_{T(R) q} \sin(\gamma_{T(R) q})
$
with the phase ones in the small $\gamma_{T(R) q}$ limit. Such a unitary transformation
can be achieved by~\cite{iskin05, zhao06}
\begin{align}
\mathbf{\bar{\Lambda}}_q = \frac{1}{\sqrt{2}}
\left( \begin{array}{cccc}
1 & \mathrm{i} & 0 & 0 \\ 
1 & -\mathrm{i} & 0 & 0 \\
0 & 0 & 1 & \mathrm{i} \\
0 & 0 & 1 & -\mathrm{i}
\end{array} \right)
\left( \begin{array}{c}
\lambda_{T q} \\ \theta_{T q} \\ \lambda_{R q} \\ \theta_{R q}
\end{array} \right),
\end{align}
where $\lambda_{T(R) q}$ and $\theta_{T(R) q}$ are real functions. 
Furthermore, assuming $\lambda_{T(R), -q} = \lambda_{T(R) q}^*$ and 
$\theta_{T(R), -q} = \theta_{T(R) q}^*$, we finally obtain the desired action
\begin{widetext}
\begin{align}
\label{eqn:S2}
\mathcal{S}_2 = \frac{1}{2T} \sum_q
\left( \lambda_{T q}^* \, \theta_{T q}^* \, \lambda_{R q}^* \, \theta_{R q}^* \right)
\left( \begin{array}{cc|cc}
T_{q,E}^{11}+T_{q}^{12} & \mathrm{i}T_{q,O}^{11} & C_{q,E}^{11}+C_{q}^{12} & \mathrm{i}C_{q,O}^{11} \\
- \mathrm{i}T_{q,O}^{11} & T_{q,E}^{11}-T_{q}^{12} & - \mathrm{i}C_{q,O}^{11} & C_{q,E}^{11}-C_{q}^{12} \\
\hline
C_{q,E}^{11*}+C_{q}^{12*} & \mathrm{i}C_{q,O}^{11*} & R_{q,E}^{11}+R_{q}^{12} & \mathrm{i}R_{q,O}^{11} \\
- \mathrm{i}C_{q,O}^{11*} & C_{q,E}^{11*}-C_{q}^{12*} & - \mathrm{i}R_{q,O}^{11} & R_{q,E}^{11}-R_{q}^{12}
\end{array} \right)
\left( \begin{array}{c}
\lambda_{T q} \\ \theta_{T q} \\ \lambda_{R q} \\ \theta_{R q}
\end{array} \right).
\end{align}
\end{widetext}
Here, we split the following matrix elements $T_q^{11}$, $R_q^{11}$ 
and $C_q^{11}$ into two in terms of an even and an odd function in 
$\mathrm{i}\nu_n$, e.g., such that
$
T_q^{11} = T_{q,E}^{11} + T_{q,O}^{11}
$ 
where
$
T_{q,E}^{11} = (T_{q}^{11} + T_{q}^{22})/2
$
is the even and
$
T_{q,O}^{11} = (T_{q}^{11} - T_{q}^{22})/2
$
is the odd part. 

Having derived the effective-Gaussian action, next we are ready to analyze 
it in detail, and extract the collective modes of the system.

\subsection{Collective excitations at $T = 0$}
\label{sec:ce0}

The dispersions $\omega_\mathbf{q}$ for the collective modes are determined 
by the poles of the propagator matrix $\mathbf{M}_q^{-1}$ for the pair fluctuation 
fields, by setting $\det \mathbf{M}_q = 0$ after an analytic continuation 
$\mathrm{i}\nu_n \to \omega + i0^+$ to the real axis. 
Since the quasiparticle-quasihole terms with the prefactor $(f-f')$ have the 
usual Landau singularity for $q \to (\mathbf{0}, 0)$ causing the collective 
modes to decay into the two-quasiparticle continuum, a small $q$ expansion 
is possible only in two cases: (i) just below the critical SF transition 
temperature provided that $\Delta_0 \to 0 \ll |\omega|$, 
and (ii) at $T = 0$ provided that $|\omega| \ll \min (E+E')$. 
In this work, we are interested in the latter case, and by setting $T = 0$ in 
Eqs.~(\ref{eqn:r1}) and~(\ref{eqn:r2}), we find
\begin{align}
\label{eqn:TqE11}
T_{q,E}^{11} &= \frac{1}{U} + \frac{1}{2N_l} \sum_{ss' \mathbf{k}} 
\frac{(\xi \xi' + E E')(E+E')} {2E E'[(\mathrm{i}\nu_n)^2 - (E+E')^2]} \nonumber\\
&\,\,\,\,\,\,\,\,\,\,\,\,\,\,\,\,\,\,\,\, \times ( 1 + d_x d_x' + d_y d_y' + d_z d_z' ), \\
\label{eqn:TqO11}
T_{q,O}^{11} &= \frac{1}{2N_l} \sum_{ss' \mathbf{k}} \frac{(\xi E' + E \xi')\mathrm{i}\nu_n}
{2E E'[(\mathrm{i}\nu_n)^2 - (E+E')^2]} \nonumber\\
&\,\,\,\,\,\,\,\,\,\,\,\,\,\,\,\,\,\,\,\, \times ( 1 + d_x d_x' + d_y d_y' + d_z d_z' ), \\
\label{eqn:Tq12}
T_{q}^{12} &= -\frac{1}{2N_l} \sum_{ss' \mathbf{k}} \frac{\Delta_0^2(E+E')}
{2E E'[(\mathrm{i}\nu_n)^2 - (E+E')^2]} \nonumber\\
&\,\,\,\,\,\,\,\,\,\,\,\,\,\,\,\,\,\,\,\, \times ( 1 + d_x d_x' + d_y d_y' + d_z d_z' ).
\end{align}
The matrix elements of $\mathbf{R}_{q}$ and $\mathbf{C}_{q}$ sectors 
have similar forms except for the multiplying factors in the second 
lines~\cite{mfactors}.

We note that, in the limit when $\mathbf{q} \to \mathbf{0}$, while the 
multiplying factor for $\mathbf{C}_q$ directly vanishes, i.e.,
$ 
d_z + d_z' - \textrm{i} (d_x d_y' - d_y d_x') \to 0,
$
for the honeycomb lattice, these terms sum to 0 for the Mielke lattice as 
$d_\mathbf{k}^z$ is an odd function of both $k_x$ and $k_y$, suggesting 
that the total and relative fluctuations are always uncoupled. In addition,
in the limit when $\mathbf{q} \to \mathbf{0}$ and $\mathrm{i}\nu_n \to 0$,
we find
$
T_{q,E}^{11} - T_{q}^{12} \to 1/U - \sum_\mathbf{k} 1/(2N_lE_{s\mathbf{k}}) = 0
$
due to the saddle-point condition, suggesting that the total phase mode 
is gapless at $\mathbf{q} = \mathbf{0}$ and we identify it as a Goldstone 
mode. Similarly, in the limit when $\mathbf{q} \to \mathbf{0}$ and 
$|\mathrm{i}\nu_n| \to 2\Delta_0$, we find
$
T_{q,E}^{11} + T_{q}^{12} \to 0
$
due again to the saddle-point condition, suggesting that the total amplitude 
mode is gapped with $2\Delta_0$ (i.e., this holds only in the weakly-interacting 
BCS limit for which the amplitude and phase fields are weakly coupled 
thanks to the negligible contribution from $T_{q,O}^{11}$) at 
$\mathbf{q} = \mathbf{0}$ and we identify it as the so-called Higgs mode. 
Therefore, we conclude that the $\mathbf{q} \to \mathbf{0}$ limit is 
consistent with our physical intuition.

We are also aware of several numerical works where the collective 
excitations of a BCS SF are analyzed on the two-dimensional honeycomb 
lattice~\cite{zhao06, tsuchiya13, zhang17}, and next we check the consistency 
of our generic results with those of the honeycomb literature.

\subsubsection{Zhao and Paramekanti's work}
\label{sec:zp}

As a first benchmark, we consider the static limit when $\mathrm{i}\nu_n \to 0$,
for which all of the matrix elements that couple amplitude and phase 
fields go to zero, i.e., 
$
\{T_{q,O}^{11}, T_{q,O}^{11}, C_{q,O}^{11}\} \to 0,
$
making the analysis of $\mathcal{S}_2$ a much simpler task. 
For instance, the phase fluctuations are described purely by the
following action
\begin{align}
\label{eqn:ap.p}
\left( \theta_{T q}^* \, \theta_{R q}^* \right)
\left( \begin{array}{cc}
T_{q,E}^{11}-T_{q}^{12} & C_{q,E}^{11}-C_{q}^{12} \\
C_{q,E}^{11*}-C_{q}^{12*} & R_{q,E}^{11}-R_{q}^{12}
\end{array} \right)
\left( \begin{array}{c}
\theta_{T q} \\ \theta_{R q}
\end{array} \right).
\end{align}
In order to confirm that Eq.~(\ref{eqn:ap.p}) reproduces the results of 
Ref.~\cite{zhao06}, one first needs to reexpress their Eq.~(4) in terms 
of the total and relative phases, and then match their matrix elements 
$u_\mathbf{q}$ and $v_\mathbf{q}$ in such a way that
$
u_\mathbf{q} + \mathrm{Re} [v_\mathbf{q}] 
=\Delta_0^2 (T_{\mathbf{q},E}^{11}-T_{\mathbf{q}}^{12}),
$
$
u_\mathbf{q} - \mathrm{Re} [v_\mathbf{q}] 
= \Delta_0^2 (R_{\mathbf{q},E}^{11}-R_{\mathbf{q}}^{12}),
$
and
$
\mathrm{Im} [v_\mathbf{q}] 
= -\Delta_0^2 (C_{\mathbf{q},E}^{11}-C_{\mathbf{q}}^{12}).
$
The origin of the prefactor $\Delta_0^2$ is due to the difference in the 
definitions of the fluctuations fields, i.e., they substitute
$
\Lambda_{Sq} = \Delta_0(\lambda_{Sq} + i\theta_{Sq}).
$
We note that since $d_\mathbf{k}^z = 0$ for the honeycomb model, their
$
\gamma_\mathbf{k} = d_\mathbf{k}^x -\mathrm{i} d_\mathbf{k}^y
$
leads to
$
\gamma_\mathbf{k}^* \gamma_\mathbf{k+q} = 
d_\mathbf{k}^x d_\mathbf{k+q}^x + d_\mathbf{k}^y d_\mathbf{k+q}^y
+\mathrm{i} (d_\mathbf{k}^x d_\mathbf{k+q}^y-d_\mathbf{k}^y d_\mathbf{k+q}^x),
$
and this expression corresponds to
$
[d_x d_x' + d_y d_y' + \mathrm{i} (d_x d_y' - d_y d_x')]/(ss')
$
in our notation. However, we note that there must be a typo in their Eq.~(5), 
and the first term must read as $\Delta_0^2/U$ instead of $2\Delta_0^2/U$. 
This is also evident from the discussion given below Eq.~(\ref{eqn:Tq12}), 
i.e., noting that $\mathrm{Im} [v_\mathbf{q}] \to 0$ in the 
$\mathbf{q} \to \mathbf{0}$ limit,
$
u_\mathbf{q} + \mathrm{Re} [v_\mathbf{q}] 
$
must also vanish in order to recover the total phase mode as a gapless 
Goldstone one. For completeness, the amplitude fluctuations are 
described purely by the following action
\begin{align}
\label{eqn:ap.a}
\left( \lambda_{T q}^* \, \lambda_{R q}^* \right)
\left( \begin{array}{cc}
T_{q,E}^{11}+T_{q}^{12} & C_{q,E}^{11}+C_{q}^{12} \\
C_{q,E}^{11*}+C_{q}^{12*} & R_{q,E}^{11}+R_{q}^{12}
\end{array} \right)
\left( \begin{array}{c}
\lambda_{T q} \\ \lambda_{R q}
\end{array} \right)
\end{align}
in the static limit.

\subsubsection{Tsuchiya, Ganesh, and Nikuni's work}
\label{sec:tgn}

As a second benchmark, we consider a two-band lattice whose energy bands 
are completely symmetric around the zero energy, i.e., 
$\xi_{s\mathbf{k}} = -\xi_{-s,\mathbf{k}}$,
which requires that $\mu = 0$ and $\epsilon_\mathbf{k} = 0$. For instance,
this particular discussion is relevant in the context of a pair of Dirac cones 
at half filling.
When this is the case, by setting $\xi_{s\mathbf{k}} = s d_\mathbf{k}$ and 
$E_{s\mathbf{k}} = \sqrt{d_\mathbf{k}^2 + \Delta_0^2} = E_\mathbf{k}$ in 
Eqs.~(\ref{eqn:TqE11})-(\ref{eqn:Tq12}), we find
\begin{align}
T_{q,E}^{11} &= \frac{1}{U} + \frac{1}{N_l} \sum_{\mathbf{k}} 
\frac{E_\mathbf{k}+E_\mathbf{k+q}} 
{E_\mathbf{k} E_\mathbf{k+q}[(\mathrm{i}\nu_n)^2 - (E_\mathbf{k}+E_\mathbf{k+q})^2]} \nonumber \\
& \times (E_\mathbf{k} E_\mathbf{k+q} + d_\mathbf{k}^x d_\mathbf{k+q}^x + d_\mathbf{k}^y d_\mathbf{k+q}^y + d_\mathbf{k}^z d_\mathbf{k+q}^z), \\
R_{q,E}^{11} &= \frac{1}{U} + \frac{1}{N_l} \sum_{\mathbf{k}} 
\frac{E_\mathbf{k}+E_\mathbf{k+q}} 
{E_\mathbf{k} E_\mathbf{k+q}[(\mathrm{i}\nu_n)^2 - (E_\mathbf{k}+E_\mathbf{k+q})^2]} \nonumber \\
&\times (E_\mathbf{k} E_\mathbf{k+q} - d_\mathbf{k}^x d_\mathbf{k+q}^x - d_\mathbf{k}^y d_\mathbf{k+q}^y + d_\mathbf{k}^z d_\mathbf{k+q}^z), \\
T_{q}^{12} &= \frac{-1}{N_l} \sum_{\mathbf{k}} \frac{\Delta_0^2(E_\mathbf{k}+E_\mathbf{k+q})}
{E_\mathbf{k} E_\mathbf{k+q} [(\mathrm{i}\nu_n)^2 - (E_\mathbf{k}+E_\mathbf{k+q})^2]}, \\
C_{q,E}^{11} &= \frac{-\mathrm{i}}{N_l} \sum_{\mathbf{k}} 
\frac{(E_\mathbf{k}+E_\mathbf{k+q}) (d_\mathbf{k}^x d_\mathbf{k+q}^y - d_\mathbf{k}^y d_\mathbf{k+q}^x)} 
{E_\mathbf{k} E_\mathbf{k+q}[(\mathrm{i}\nu_n)^2 - (E_\mathbf{k}+E_\mathbf{k+q})^2]}, \\
C_{q,O}^{11} &= \frac{1}{N_l} \sum_{\mathbf{k}} 
\frac{(d_\mathbf{k}^z E_\mathbf{k+q} + E_\mathbf{k} d_\mathbf{k+q}^z) \mathrm{i}\nu_n} 
{E_\mathbf{k} E_\mathbf{k+q}[(\mathrm{i}\nu_n)^2 - (E_\mathbf{k}+E_\mathbf{k+q})^2]}.
\end{align}
The remaining terms are such that $R_{q}^{12} = T_{q}^{12}$, and 
$T_{q,O}^{11} = R_{q,O}^{11}  = C_{q}^{12} = 0$. We also note that 
the saddle-point condition Eq.~(\ref{eqn:op}) becomes 
$
1/U = \sum_\mathbf{k} 1/(N_l E_\mathbf{k}).
$
Since $C_{q,O}^{11}$ vanishes for the honeycomb lattice, and it sums 
to $0$ for $d_\mathbf{k}^z$ that is odd in $k_x$ or $k_y$, we find for 
these cases that the amplitude and phase fields are 
completely decoupled, i.e., they are purely described by Eqs.~(\ref{eqn:ap.p})
and~(\ref{eqn:ap.a}). Setting the corresponding determinants to 0, 
we find
\begin{align}
&\bigg\lbrace \frac{1}{U} + \sum_{\mathbf{k}} 
\frac{(E_\mathbf{k}+E_\mathbf{k+q}) (E_\mathbf{k} E_\mathbf{k+q}  \pm \Delta_0^2 + d_\mathbf{k}^z d_\mathbf{k+q}^z)} 
{N_l E_\mathbf{k} E_\mathbf{k+q}[(\mathrm{i}\nu_n)^2 - (E_\mathbf{k}+E_\mathbf{k+q})^2]} \bigg\rbrace^2 \nonumber \\
& \,\,\,\,= \bigg\lbrace \frac{1}{N_l}  \sum_{\mathbf{k}}
\frac{(E_\mathbf{k}+E_\mathbf{k+q}) (d_\mathbf{k}^x d_\mathbf{k+q}^x + d_\mathbf{k}^y d_\mathbf{k+q}^y)} 
{E_\mathbf{k} E_\mathbf{k+q}[(\mathrm{i}\nu_n)^2 - (E_\mathbf{k}+E_\mathbf{k+q})^2]} \bigg\rbrace^2 \nonumber \\
&\,\,\,\,\,+ 
\bigg\lbrace \frac{1}{N_l}  \sum_{\mathbf{k}}
\frac{(E_\mathbf{k}+E_\mathbf{k+q}) (d_\mathbf{k}^x d_\mathbf{k+q}^y - d_\mathbf{k}^y d_\mathbf{k+q}^x)} 
{E_\mathbf{k} E_\mathbf{k+q}[(\mathrm{i}\nu_n)^2 - (E_\mathbf{k}+E_\mathbf{k+q})^2]} \bigg\rbrace^2,
\label{eqn:tgn}
\end{align}
for the poles of the propagator matrices given in Eqs.~(\ref{eqn:ap.p}) 
and~(\ref{eqn:ap.a}), where $\pm$ is for the phase/amplitude modes. 
Note that since $C_{q,E}^{11} \to 0$ in the limit when $\mathbf{q} \to \mathbf{0}$, 
the total and relative fields are not coupled, leading to a gapless 
Goldstone mode and a gapped Leggett mode as discussed below and 
in Sec.~\ref{sec:prf}.

In the honeycomb case, Eq.~(\ref{eqn:tgn}) is in somewhat agreement 
with Ref.~\cite{tsuchiya13}, i.e., our $\pm$ results are similar to their 
expressions Eqs.~(12) and~(11), respectively, when their $F = 0$. 
This discrepancy is amusing given that the collective modes for the 
usual one-band models that are found from the Gaussian fluctuations 
and random-phase approximation are known to be consistent with 
each other. Furthermore, they conclude that the Goldstone and Leggett 
modes are both gapless and degenerate at $\mathbf{q} = \mathbf{0}$.

When we set $\mathbf{q} = \mathbf{0}$ in Eq.~(\ref{eqn:tgn}), we find two 
solutions for the phase modes and two solutions for the amplitude ones, 
which can be written, respectively, as
\begin{align}
\label{eqn:tgntp}
0 &= \frac{1}{N_l} \sum_\mathbf{k} \frac{(\mathrm{i}\nu_n)^2}
{E_\mathbf{k} [(\mathrm{i}\nu_n)^2 - 4E_\mathbf{k}^2] }, \\
\label{eqn:tgnrp}
0 &= \frac{1}{N_l} \sum_\mathbf{k} \frac{(\mathrm{i}\nu_n)^2 - 4d_\mathbf{k}^2 + 4(d_\mathbf{k}^z)^2}
{E_\mathbf{k} [(\mathrm{i}\nu_n)^2 - 4E_\mathbf{k}^2] }, \\
\label{eqn:tgnta}
0 &= \frac{1}{N_l} \sum_\mathbf{k} \frac{(\mathrm{i}\nu_n)^2 - 4\Delta_0^2}
{E_\mathbf{k} [(\mathrm{i}\nu_n)^2 - 4E_\mathbf{k}^2] }, \\
\label{eqn:tgnra}
0 &= \frac{1}{N_l} \sum_\mathbf{k} \frac{(\mathrm{i}\nu_n)^2 - 4d_\mathbf{k}^2 + 4(d_\mathbf{k}^z)^2 - 4\Delta_0^2}
{E_\mathbf{k} [(\mathrm{i}\nu_n)^2 - 4E_\mathbf{k}^2] }.
\end{align}
Here, Eq.~(\ref{eqn:tgntp}) suggests that the total phase (Goldstone) mode is
gapless when $\mathrm{i} \nu_n \to 0$, and Eq.~(\ref{eqn:tgnta}) suggests
that the total amplitude mode is gapless when $|\mathrm{i} \nu_n| \to 2\Delta_0$.
In addition, the relative phase (Leggett) mode is gapped as Eq.~(\ref{eqn:tgnrp}) 
is not satisfied for $\mathrm{i} \nu_n \to 0$, and assuming 
$|\mathrm{i}\nu_n| \ll \min (2E_\mathbf{k}) = 2\Delta_0$, its finite frequency 
is determined by
$
(\mathrm{i}\nu_n)^2 = \{\sum_\mathbf{k} [d_\mathbf{k}^2-(d_\mathbf{k}^z)^2]/E_\mathbf{k}^3\} 
/ \{\sum_\mathbf{k} [\Delta_0^2+(d_\mathbf{k}^z)^2] /(4E_\mathbf{k}^5)\}.
$
This is in agreement with the low-frequency expansion that is presented 
in Sec.~\ref{sec:prf}, where $\omega_L^2 \to \widetilde{P}/\widetilde{R}$
at $\mathbf{q} = \mathbf{0}$. 
Applying a similar analysis to Eq.~(\ref{eqn:tgnra}), we find that the finite 
frequency of the relative amplitude mode is determined by
$
(\mathrm{i}\nu_n)^2 = \{\sum_\mathbf{k} [E_\mathbf{k}^2-(d_\mathbf{k}^z)^2]/E_\mathbf{k}^3\} 
/ [\sum_\mathbf{k} (d_\mathbf{k}^z)^2 /(4E_\mathbf{k}^5)],
$
and it is much larger than $2\Delta_0$. This clearly suggests that this mode 
is always damped, and it decays into the two-quasiparticle continuum. 
For instance, in the strong-coupling BEC limit when 
$\Delta_0 \gg \max{d_\mathbf{k}}$, these frequencies can be 
approximated by
$
(\mathrm{i}\nu_n)^2 = (8/N_l)\sum_\mathbf{k} [d_\mathbf{k}^2-(d_\mathbf{k}^z)^2]
$
for the undamped Leggett mode, and by
$
(\mathrm{i}\nu_n)^2 = 2 N_l \Delta_0^4/ \sum_\mathbf{k} (d_\mathbf{k}^z)^2
$
for the damped relative amplitude one. The former result is consistent 
with the recent literature, where undamped Leggett modes are found 
for sufficiently strong interactions away from the weak-coupling 
BCS limit~\cite{zhao06, zhang17}.
As a final remark, setting $d_\mathbf{k}^z = 0$ in Eq.~(\ref{eqn:tgnra})
for the honeycomb case, we simply find $0 = 1/U$, suggesting that 
the relative amplitude branch disappears at $\mathbf{q} = \mathbf{0}$.

\section{Geometric interpretation}
\label{sec:gi}

As discussed in Sec.~\ref{sec:ce0}, the total and relative fluctuations turn
out to be uncoupled from each other in the limit when $\mathbf{q} \to \mathbf{0}$. 
Next we consider this limit, and discuss purely total and purely relative 
fluctuations in detail due to their analytical simplicity.

\subsection{Purely total fluctuations}
\label{sec:ptf}

For this purpose, it is sufficient to take into account the following terms 
in the small $\mathbf{q}$ and $\omega$ expansions:
$
T_{q,E}^{11}+T_q^{12} = A + \sum_{ij} C_{ij} q_i q_j - D\omega^2 + \cdots;
$
$
T_{q,E}^{11}-T_q^{12} = \sum_{ij} Q_{ij} q_i q_j - R\omega^2 + \cdots;
$
and
$
T_{q,O}^{11} = -B\omega + \cdots.
$
Since $B \ne 0$ in general, it couples the total phase and total amplitude 
fields, and therefore, we derive a total phase-(amplitude)-only action
 by integrating out the total amplitude (phase) fields. This leads to a 
 phonon-like gapless in-phase (Goldstone) mode and an exciton-like 
 gapped amplitude (Higgs) mode~\cite{totalnote}
\begin{align}
\label{eqn:wGq}
\omega_{G\mathbf{q}}^2 &= \sum_{ij} \frac{Q_{ij}}{R+ B^2/A} q_i q_j, \\
\label{eqn:wHq}
\omega_{H\mathbf{q}}^2 &= \frac{A + B^2/R}{D} 
+ \sum_{ij} \left( \frac{C_{ij}}{D} + \frac{B^2Q_{ij}/R}{B^2+ A R} \right) q_i q_j.
\end{align}
Here, the nonkinetic coefficients are given by
$
A = \sum_{s \mathbf{k}} \Delta_0^2/(2N_lE_{s\mathbf{k}}^3),
$
$
B = \sum_{s \mathbf{k}} \xi_{s\mathbf{k}}/(4N_lE_{s\mathbf{k}}^3),
$
$
D = \sum_{s \mathbf{k}} \xi_{s\mathbf{k}}^2/(8N_lE_{s\mathbf{k}}^5),
$
and
$
R = \sum_{s \mathbf{k}} 1/(8N_lE_{s\mathbf{k}}^3).
$
We note that these expressions are simply summations over their 
conventional counterparts for the usual one-band problem, i.e., 
they are due entirely to intraband mechanisms. 

On the other hand, the kinetic coefficients have a tensor structure, and they 
consist of both an intraband and an interband contribution in such a way that
$
C_{ij} = C_{ij}^\mathrm{intra} + C_{ij}^\mathrm{inter}
$
and
$
Q_{ij} = Q_{ij}^\mathrm{intra} + Q_{ij}^\mathrm{inter}.
$
A compact way to express these coefficients are
\begin{align}
\label{eqn:Cintra}
C_{ij}^\mathrm{intra} &= \frac{1}{N_l}\sum_{s \mathbf{k}} \frac{1}{8 E_{s\mathbf{k}}^3} 
\left( 1 - \frac{5\Delta_0^2 \xi_{s\mathbf{k}}^2}{E_{s\mathbf{k}}^4} \right)
\frac{\partial \xi_{s\mathbf{k}}}{\partial k_i} \frac{\partial \xi_{s\mathbf{k}}}{\partial k_j}, \\
\label{eqn:Qintra}
Q_{ij}^\mathrm{intra} &= \frac{1}{N_l}\sum_{s \mathbf{k}} \frac{1}{8 E_{s\mathbf{k}}^3}
\frac{\partial \xi_{s\mathbf{k}}}{\partial k_i} \frac{\partial \xi_{s\mathbf{k}}}{\partial k_j}, \\
\label{eqn:Cinter}
C_{ij}^\mathrm{inter} &= -  \frac{1}{N_l} \sum_{s \mathbf{k}} \frac{d_\mathbf{k}}{4 s\xi_\mathbf{k} E_{s\mathbf{k}}}
\left( 1 + \frac{2\Delta_0^2}{d_\mathbf{k}^2} \right) g_\mathbf{k}^{ij}, \\
\label{eqn:Qinter}
Q_{ij}^\mathrm{inter} &= - \frac{1}{N_l} \sum_{s \mathbf{k}}
\frac{d_\mathbf{k}}{4 s\xi_\mathbf{k} E_{s\mathbf{k}}} g_\mathbf{k}^{ij}.
\end{align}
We again note that while Eqs.~(\ref{eqn:Cintra}) and~(\ref{eqn:Qintra}) 
can be expressed as a sum over their conventional conunterparts, 
Eqs.~(\ref{eqn:Cinter}) and~(\ref{eqn:Qinter}) do not have counterparts 
in the usual one-band problem. It turns out that the interband contributions 
are controlled by the quantum metric tensor $g_{\mathbf{k}}^{ij}$ of the 
underlying quantum states in $\mathbf{k}$ space~\cite{provost80,berry89,resta11}. 
For our generic two-band lattice model, the quantum metric tensor of the
Bloch states can be written as
$
2g_{\mathbf{k}}^{ij} = -\widehat{\mathbf{d}}_\mathbf{k} \cdot 
\partial^2 \widehat{\mathbf{d}}_\mathbf{k} / (\partial k_i \partial k_j)
$
or equivalently
$
2g_{\mathbf{k}}^{ij} = (\partial \widehat{\mathbf{d}}_\mathbf{k}/\partial k_i) 
\cdot (\partial \widehat{\mathbf{d}}_\mathbf{k}/\partial k_j).
$
Alternatively, it can be expressed as
\begin{align}
g_\mathbf{k}^{ij} = \frac{1}{2d_\mathbf{k}^2} 
\sum_{\ell = (x,y,z)} \frac{\partial d_\mathbf{k}^\ell}{\partial k_i} 
\frac{\partial d_\mathbf{k}^\ell}{\partial k_j} 
- 
\frac{1}{2d_\mathbf{k}^2}
\frac{\partial d_\mathbf{k}}{\partial k_i} \frac{\partial d_\mathbf{k}}{\partial k_j},
\end{align}
without the loss of generality.

\subsection{Purely relative fluctuations}
\label{sec:prf}

Similar to Sec.~\ref{sec:ptf}, it may again be sufficient to take into account the 
following terms in the small $\mathbf{q}$ and $\omega$ expansions:
$
R_{q,E}^{11}+R_q^{12} = \widetilde{A} + \sum_{ij} \widetilde{C}_{ij} q_i q_j 
- \widetilde{D}\omega^2 + \cdots;
$
$
R_{q,E}^{11}-R_q^{12} = \widetilde{P} + \sum_{ij} \widetilde{Q}_{ij} q_i q_j - \widetilde{R}\omega^2 + \cdots;
$
and
$
R_{q,O}^{11} = -\widetilde{B}\omega + \cdots.
$
None of these expansion coefficients have a conventional counterpart in 
the usual one-band model. For instance, the nonkinetic coefficients are 
given by
\begin{align}
\widetilde{A} (\widetilde{P})  &= \frac{1}{U} -\frac{1}{N_l} \sum_{ss'\mathbf{k}}
\frac{\xi_{s\mathbf{k}} \xi_{s' \mathbf{k}} + E_{s\mathbf{k}} E_{s'\mathbf{k}} \mp \Delta_0^2}
{4E_{s\mathbf{k}} E_{s'\mathbf{k}} (E_{s\mathbf{k}} + E_{s'\mathbf{k}})} x_{ss'}^\mathbf{k}, \\
\widetilde{D} (\widetilde{R}) &= \frac{1}{N_l} \sum_{ss'\mathbf{k}}
\frac{\xi_{s\mathbf{k}} \xi_{s' \mathbf{k}} + E_{s\mathbf{k}} E_{s'\mathbf{k}} \mp \Delta_0^2}
{4E_{s\mathbf{k}} E_{s'\mathbf{k}} (E_{s\mathbf{k}} + E_{s'\mathbf{k}})^3} x_{ss'}^\mathbf{k}, \\
\widetilde{B} &= \frac{1}{N_l} \sum_{ss'\mathbf{k}}
\frac{\xi_{s\mathbf{k}} E_{s' \mathbf{k}} + E_{s\mathbf{k}} \xi_{s'\mathbf{k}}}
{4E_{s\mathbf{k}} E_{s'\mathbf{k}} (E_{s\mathbf{k}} + E_{s'\mathbf{k}})^2} x_{ss'}^\mathbf{k}, 
\end{align}
where we define
$
x_{ss'}^\mathbf{k} = 1 - ss' [ (d_\mathbf{k}^x)^2 + (d_\mathbf{k}^y)^2 - 
(d_\mathbf{k}^z)^2 ] / d_\mathbf{k}^2.
$
The kinetic coefficients $\widetilde{C}_{ij}$ and $\widetilde{Q}_{ij}$ are
more involved and not presented here. 

This expansion suggests that the Leggett mode is gapped as long as 
$\widetilde{P} \ne 0$, and its finite frequency is determined by
$
\omega_L^2 = \widetilde{P}/\widetilde{R},
$
when the coupling between the relative phase and relative amplitude 
fields is negligible. Here we note an intuitive result that 
$\widetilde{P} = 0$ when the two bands are identical, i.e., when the 
sublattice-coupling field $\mathbf{d}_\mathbf{k} = 0$ vanishes so that 
$
\xi_{s\mathbf{k}} = \xi_{-s,\mathbf{k}} = \epsilon_\mathbf{k} - \mu.
$
In addition, in the strong-coupling BEC limit when 
$\Delta_0 \gg \max |\epsilon_{s\mathbf{k}}|$, we note that $\widetilde{P} \to 0$ 
as well. This is because since $\mu \lessgtr 0$ and 
$|\mu| \gg \max |\epsilon_{s\mathbf{k}}|$ in the dilute limit of particles/holes 
when $0.5 - N_0/N_l \approx \pm 0.5$, and $|\mu| \approx 0$ 
around half filling when $N_0/N_l \approx 0.5$, one can substitute
$\xi_{s\mathbf{k}} \to -\mu$ and 
$E_{s\mathbf{k}} \to \sqrt{\mu^2 + \Delta_0^2}$.
Thus, we conclude that the Leggett mode becomes undamped for 
sufficiently strong interactions with a negligibly smaller gap in the 
strong-coupling limit. This result is also intuitive given that the 
sublattice structure of the non-interacting particles should not 
play a primary role in the regime of tightly-bound molecules.

Since $\widetilde{B} \ne 0$ in most cases, we derive a 
relative phase-(amplitude)-only action by integrating out the relative amplitude 
(phase) fields. This leads to an exciton-like out-of-phase (Leggett) mode
and an exciton-like higher-energy amplitude (Higgs) mode~\cite{relativenote}
\begin{align}
\omega_{L(H)\mathbf{q}}^2 &= \frac{\widetilde{B}^2 + \widetilde{A}\widetilde{R} 
+ \widetilde{P} \widetilde{D} \mp \widetilde{W}}
{2 \widetilde{D} \widetilde{R}} \nonumber \\
&+ \sum_{ij} \left[
\frac{\widetilde{C}_{ij}}{2\widetilde{D}} \left( 1 \mp \frac{\widetilde{B}^2 + \widetilde{A}\widetilde{R} 
- \widetilde{P} \widetilde{D}}{\widetilde{W}} \right) \right. \nonumber \\
&\,\,\,\,\,\,\,\,\,\,\,\,\,+ 
\left. \frac{\widetilde{Q}_{ij}}{2\widetilde{R}} \left(1 \mp \frac{\widetilde{B}^2 - \widetilde{A}\widetilde{R} 
+ \widetilde{P} \widetilde{D}}{\widetilde{W}}\right) 
\right] q_i q_j,
\label{eqn:rmodes}
\end{align}
where we define
$
\widetilde{W} = [(\widetilde{B}^2 + \widetilde{A}\widetilde{R} + 
\widetilde{P} \widetilde{D})^2 - 4 \widetilde{A} \widetilde{P}\widetilde{D} \widetilde{R}]^{1/2}.
$
Here, the leading nonzero contribution to $\beta_5$ is approximated by 
$\widetilde{D} \widetilde{R}$~\cite{relativenote}, and it must be replaced with 
the proper factor coming from the higher-order expansion coefficients in those 
exceptional cases when $\widetilde{D} = 0$. One such example is the honeycomb 
lattice that is considered in Sec.~\ref{sec:tgn}, for which case we find 
$\widetilde{B} = 0$ and set 
$
\widetilde{W} = \widetilde{A}\widetilde{R} - \widetilde{P}\widetilde{D},
$
where
$
\widetilde{A} = 1/U = \sum_\mathbf{k} 1/(N_l E_\mathbf{k}),
$
$
\widetilde{P} = \sum_\mathbf{k} d_\mathbf{k}^2/(N_l E_\mathbf{k}^3),
$
$
\widetilde{R} = \sum_\mathbf{k} \Delta_0^2/(4 N_l E_\mathbf{k}^5),
$
and 
$
\widetilde{D} = 0.
$

\subsection{SF phase stiffness tensor}
\label{sec:spst}

At $T = 0$, we verify that the SF phase stiffness tensor $\mathcal{D}_{ij}$ 
is directly proportional to the kinetic coefficient $Q_{ij}$ of the total phase 
fluctuations, i.e.,  
$
\mathcal{D}_{ij} = 8 N_l (\Delta_0^2/\mathcal{A}) Q_{ij}
$
with $\mathcal{A}$ the area of the lattice, in such a way 
that~\cite{liang17, iskin19a, iskin19b}
\begin{align}
\label{eqn:Dconv}
\mathcal{D}_{ij}^\textrm{conv} &= \frac{\Delta_0^2}{\mathcal{A}} \sum_{s \mathbf{k}} 
\frac{1}{E_{s\mathbf{k}}^3}
\frac{\partial \xi_{s\mathbf{k}}}{\partial k_i} \frac{\partial \xi_{s\mathbf{k}}}{\partial k_j}, \\
\label{eqn:Dgeom}
\mathcal{D}_{ij}^\textrm{geom} &= - \frac{2\Delta_0^2}{\mathcal{A}} \sum_{s \mathbf{k}}
\frac{d_\mathbf{k}}{s\xi_\mathbf{k} E_{s\mathbf{k}}} g_\mathbf{k}^{ij}.
\end{align}
In addition to those given in Sec.~\ref{sec:zp} and~\ref{sec:tgn}, this 
association may be considered as a third benchmark for the consistency 
of our results with the recent literature. The direct link between the 
quantum metric tensor and the SF stiffness tensor is relatively new 
in the literature~\cite{peotta15, liang17}, revealing the geometric 
origin of superconductivity in the presence of other bands. This result is 
particularly illuminating for a narrow- or flat-band superconductivity for 
which the geometric contribution clearly dominates the SF stiffness 
tensor when the conventional one is negligible. Motivated by these works, 
there have been many studies on the subject exploring a variety of 
multi-band Hamiltonians, including most recently that of the twisted bilayer 
graphene~\cite{hu19, julku19, xie19}. 

Furthermore, it has been proposed that the quantum metric tensor has a 
partial control over all those SF properties that depend explicitly on 
the effective-mass tensor of the SF carriers, i.e., of the corresponding 
(two- or many-body) bound state~\cite{iskin19a, iskin19b}. 
In the context of two-band SFs, our finding Eq.~(\ref{eqn:wGq}) for the 
velocity of the Goldstone mode is in complete agreement with our earlier 
work~\cite{iskin19c}, suggesting that an analogous contribution to 
the collective excitations must be present in many other multi-band 
systems as well.

\section{Conclusions}
\label{sec:conc}

In summary, we considered a generic lattice Hamiltonian that is suitable 
for describing a uniform BCS SF with two underlying sublattices, and then 
extracted its collective excitations from an effective action that is derived up 
to the Gaussian order in the fluctuations of the SF order parameter. 
Allowing for independent fluctuations on the two sublattices, there are 
phonon-like in-phase (Goldstone) and exciton-like out-of-phase (Leggett) 
modes in this system. While the Goldstone mode is gapless at zero 
momentum and propagating in general, the Leggett mode becomes 
undamped only with sufficiently strong interactions.
Furthermore, we showed that, in addition to the conventional contribution, 
the velocity of the Goldstone mode has a geometric contribution that is 
governed by the quantum metric tensor of the Bloch states. This suggests 
that the latter contribution dominates the velocity when the former one 
becomes negligible for a narrow- or a flat-band model. We traced the 
origin of the geometric contribution to the Goldstone mode back 
to the recent works on the geometric contribution to the SF stiffness tensor, 
and argued that these geometric effects are complementary to each other, 
i.e., they are both controlled by the effective-mass tensor of the SF carriers.
This suggests that an analogous contribution to the collective excitations 
must be present in many other multi-band systems including the twisted 
bilayer graphene~\cite{hu19, julku19, xie19}. As a further outlook, it is also 
worthwhile to study the damping of these collective excitations at finite 
temperatures~\cite{klimin19}.

\begin{acknowledgments}
This work is supported by the funding from T{\"U}B{\.I}TAK Grant No. 1001-118F359.
\end{acknowledgments}


\begin{thebibliography}{99}
\bibitem{totalnote}
For the purely total fluctuations, one needs to derive the characteristic 
equation up to fourth order in the expansion, e.g., 
$
\beta_5 \omega^4 + \sum_{ijkl} \beta_4^{ijkl} q_i q_j q_k q_l 
+ \sum_{ij} \beta_3^{ij} \omega^2 q_i q_j + \beta_2\omega^2 
+ \sum_{ij} \beta_1^{ij} q_iq_j = 0,
$
such that the Goldstone mode 
$
\omega_{G\mathbf{q}}^2 = \sum_{ij} x_G^{ij} q_i q_j
$
is determined by
$
x_G^{ij} = -\beta_1^{ij}/\beta_2,
$
and the Higgs mode 
$
\omega_{H\mathbf{q}}^2 = \omega_H^2 + \sum_{ij} x_H^{ij} q_i q_j
$
is determined by
$
\omega_H^2 = -\beta_2/\beta_5
$
and
$
x_H^{ij} = -\beta_3^{ij}/\beta_5 + \beta_1^{ij}/\beta_2.
$
In the main text, while our quadratic expansion fully determines $x_G^{ij}$, 
i.e., our Eq.~(\ref{eqn:wGq}) is exact, we neglect the additional corrections to 
$\omega_H^2$ and $x_H^{ij}$ that are coming from the higher-order terms 
to $\beta_3^{ij}$ and $\beta_5$. To be more precise, we substitute
$
\beta_1^{ij} = A Q_{ij},
$
$
\beta_2 = - AR - B^2,
$
$
\beta_3^{ij} \approx - RC_{ij} - DQ_{ij},
$
and
$
\beta_5 \approx D R.
$
\bibitem{relativenote}
For the purely relative fluctuations, one needs to derive the characteristic 
equation up to fourth order in the expansion, e.g., 
$
\beta_5 \omega^4 + \sum_{ijkl} \beta_4^{ijkl} q_i q_j q_k q_l 
+ \sum_{ij} \beta_3^{ij} \omega^2 q_i q_j + \beta_2\omega^2 
+ \sum_{ij} \beta_1^{ij} q_iq_j + \beta_0= 0,
$
such that the Leggett and Higgs modes 
$
\omega_{L(H)\mathbf{q}}^2 = \omega_{L(H)}^2 + \sum_{ij} x_{L(H)}^{ij} q_i q_j
$
are determined by
$
\omega_{L(H)} = \left(-\beta_2 \mp \sqrt{\beta_2^2 - 4\beta_0\beta_5} \right)/(2\beta_5)
$
and
$
x_{L(H)}^{ij} = [-\beta_3^{ij} \pm (\beta_2\beta_3^{ij}-2\beta_5\beta_1^{ij})
/\sqrt{\beta_2^2 - 4\beta_0\beta_5}]/(2\beta_5) .
$
Note that Ref.~\cite{totalnote} is recovered in the $\beta_0 \to 0$ limit.
In our quadratic expansion presented in the main text, we neglect the 
additional corrections to $\omega_{L(H)}^2$ and $x_{L(H)}^{ij}$ that are
coming from the higher-order terms to $\beta_3^{ij}$ and $\beta_5$.
To be more precise, we substitute
$
\beta_0 = \widetilde{A} \widetilde{P},
$
$
\beta_1^{ij} = \widetilde{A} \widetilde{Q}_{ij} + \widetilde{P} \widetilde{C}_{ij},
$
$
\beta_2 = - \widetilde{A} \widetilde{R} - \widetilde{P} \widetilde{D} - \widetilde{B}^2,
$
$
\beta_3^{ij} \approx - \widetilde{R} \widetilde{C}_{ij} - \widetilde{D} \widetilde{Q}_{ij},
$
and
$
\beta_5 \approx \widetilde{D} \widetilde{R}.
$
\end{thebibliography}
\end{document}